# High mobility back-gated InAs/GaSb double quantum well grown on GaSb substrate.

Binh-Minh Nguyen*, Wei Yi, Ramsey Noah, Jacob Thorp, Marko Sokolich*

HRL Laboratories, 3011 Malibu Canyon Rd, Malibu, CA 90265

Abstract:

We report a backgated InAs/GaSb double quantum well device grown on GaSb substrate. The use of the native substrate allows for high materials quality with electron mobility in excess of 500,000 cm$^2$/Vs at sheet charge density of 8x10$^{11}$ cm$^{-2}$ and approaching 100,000 cm$^2$/Vs near the charge neutrality point (CNP). Lattice matching between the quantum well structure and the substrate eliminates the need for a thick buffer, enabling large back gate capacitance and efficient coupling with the conduction channels in the quantum wells. As a result, quantum Hall effects are observed in both electron and hole regimes across the hybridization gap.

*Corresponding authors: mbnguyen@hrl.com and MSokolich@hrl.com

The unique band alignment between InAs/GaSb in which the conduction band edge of InAs lies below the valence band edge of GaSb offers flexible control over the electronic band structure of the material system. Thin InAs/GaSb quantum wells or superlattices behave as conventional semiconductor with an effective bandgap tunable across the entire infrared regime, whereas thicker structures enter the "semimetallic" regime since the quantized energy levels for electron and holes are close to the bulk band edges[1-5]. It is shown that in such a regime, a hybridization gap of a few meV exists, creating a medium where electrons and hole states are mixed[6-11]. In recent years, interest in such an electron-hole interaction in the InAs/GaSb structure was revived as it was proposed that the material system possessed quantum spin Hall phase.[12] While significant progress has been made with reports on observation of bulk insulating state, helical edge transport mode in the hybridization gap[13-18], the material structure and device design remain the same as they were originally published in the 1990s[19,20]. A main drawback of the structure is the use of a 7% lattice mismatch GaAs substrate that requires the use of a thick buffer for dislocation reduction. Despite the thick buffer, transport in the InAs/GaSb quantum well channels still suffers from high dislocation density and quantum well-barrier interface roughness of the same order as the quantum well width (as evidenced by surface roughness). Typically, surface roughness of the InAs/GaSb system on GaAs substrates is a few Angstroms[21,22] up to above 1nm[23], which is greater than 10% of the widths of the quantum wells (e.g. 50 Å for GaSb and 125 Å for InAs in typical double quantum well structure). An electron mobility as high as 300,000 $cm^2$/Vs has been reported, but only at high sheet charge density ($1.5 \times 10^{12}$ $cm^{-2}$) which allows sufficient screening; near the charge neutrality point, electron mobility is only 20,000 $cm^{-2}$/Vs in the same work.[17] It is still highly desirable to further enhance the material quality and interface abruptness of the InAs/GaSb structure, both of which can be evidenced by higher carrier mobility. In the absence of ionized impurities, interface roughness is a primary cause of mobility degradation in narrow InAs quantum wells.[24] Enhancing mobility should lead directly to enhancement of the resonance peak in the magnetoresistance, Rxx, observed when the Fermi level is in the hybridization gap.[25] This in turn should enable the observation of edge mode conduction with greater fidelity.

Growth of InAs/GaSb on GaSb is not new; the field of infrared detection has greatly benefited from the use of InAs/GaSb superlattices grown on commercially available, lattice matched, native GaSb substrates[26]. Defect free detectors grown on GaSb substrate exhibit orders of magnitude higher differential resistance than devices grown on GaAs substrates, facilitating recent developments in infrared imaging devices.[27-31] On the other hand, fundamental studies of quantum Hall effect continue to use InAs/GaSb grown on GaAs substrates with thick, strain-relaxed buffer layers with the disadvantages discussed above. In this letter, we demonstrate the advantage of using GaSb substrates with enhanced carrier mobilities, and efficient electrostatic coupling between the backgate and the conduction channels in the quantum Hall regime.

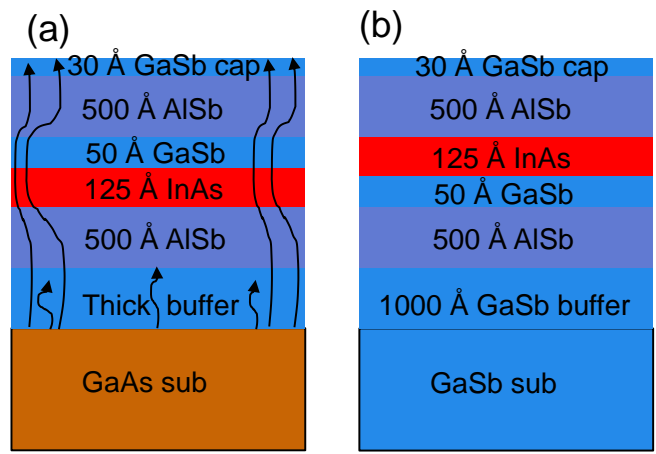

Fig1- Schematic diagram of (a) typical device structures reported in the literature (see for example Ref[14]) and (b) design used in this work. Arrowed lines in figure (a) illustrate dislocation network. Thicknesses are not to scale.

Fig 1 compares the schematic diagram of a typical device design reported in the literature and the structure used in this work. Reference material (Fig1a) was grown on GaAs substrate, on top of an AlSb nucleation layer and a few micrometer thick buffer layer. The device core consists of a GaSb hole well on top of an InAs electron well, which are sandwiched between 500 Å thick AlSb barriers. The structure is finally capped with a 30 Å thick GaSb oxidation-prevention layer. In contrast, the sequence of the quantum wells in our design (Fig1b) is inverted, namely InAs on GaSb as opposed to GaSb on InAs in published structures so that the action of the bottom gate would be similar to that of the top gate configuration previously published. The core device thicknesses remain the same as in ref[13] with 50 Å thick GaSb and 125 Å thick InAs wells and 500 Å thick AlASb barriers. The key difference between the current work and previous results is the use of a GaSb substrate, closely lattice matched to the InAs/GaSb/AlSb system instead of the 7% lattice mismatched GaAs substrate. Besides expected improvement in the InAs/GaSb layer quality, the structure eliminates the need for a thick, defective buffer which (1) still suffers from breakdown leakage through the dislocation network and (2) results in low backgate capacitance. The weak coupling between the backgate voltage and the quantum well conduction channels due to thick buffer requires large gate bias to modulate the conductance, which is compromised with how much field the gate "dielectric" bears before breakdown.  As will be shown in this letter, a non-intentionally doped GaSb buffer as thin as 1000 Å is sufficient to result in an atomically abrupt interface between the layers, while, together with the 500 Å thick AlSb barrier, can electrically isolate the InAs/GaSb quantum well channels from the GaSb substrate used as back gate contact.  A leakage current below 50 pA can be maintained across the bias range that fully enables characterization of quantum Hall effects through the electron, hole and charge neutrality regimes. The advantage of the backgated structure is the simple device processing, the epitaxial interface between the gate contact (substrate), the dielectric layer (GaSb buffer and AlSb barrier) and the conduction channels (InAs/GaSb quantum wells), which can minimize undesired hysteresis. When combined with a properly configured top-gate, this allows for independent control of individual channels, which is highly desirable in a hybridized system such as InAs/GaSb double quantum wells.

Materials used in this work were grown on an n-type Te-doped GaSb substrate using a Varian Gen II molecular beam epitaxy (MBE) system equipped with valved crackers for As, Sb and Sumo© cells for Ga and In. Growth conditions for the buffers and quantum wells were similar to previously published[32]. As grown materials were characterized using high resolution x-ray diffraction (HRXRD) and high resolution transmission electron microscopy (TEM). Despite the absence of periodical structures, oscillation fringes are clearly observed in measured HRXRD rocking curves, and match very well with simulation using PANalytical's X'Pert Epitaxy software. The complex diffraction patterns are an indication of high crystallinity, and abrupt interfaces. It is important to note that the simulated fringe structures are very sensitive to small changes (less than 10%) in individual layer thicknesses; thus X-ray simulation is a simple yet effective technique to cross-check growth rate calibrations. More precise determination of layer thicknesses was done with HRTEM, as shown in Fig 2b,c. No sign of dislocation is observed in Fig 2b, and even at lower magnification with wider field of view (not shown). High resolution TEM (Fig 2c) clearly shows atomically sharp interfaces between the core layers. Measured thicknesses are 125Å, 52 Å and 520 Å for InAs, GaSb and AlSb, respectively. Analysis using atomic force microscopy (not shown) exhibits a top surface roughness of only 1.5 Å over a 5x5mm$^2$ area scan, which is similar to typical morphology of InAs/GaSb superlattice on GaSb for infrared detector[33], and is significantly smoother than morphology of InAs/GaSb structures on GaAs as commented above.

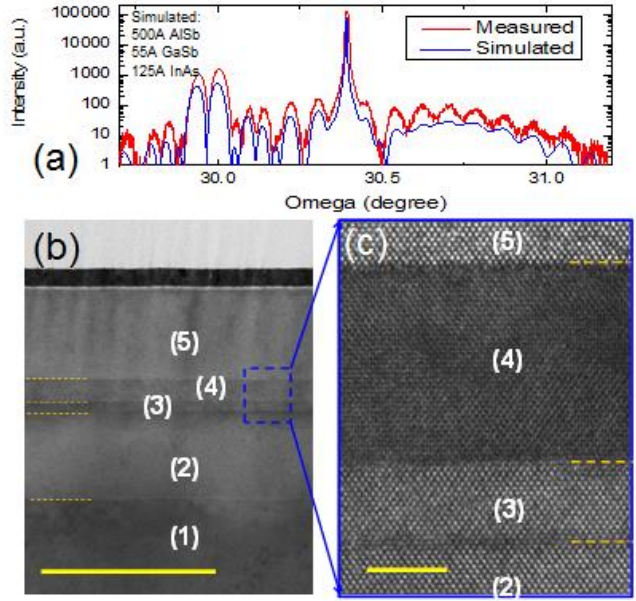

Fig2- (a) HRXRD rocking curve and simulation of the device structure, (b) TEM and (c) HRTEM image of the device cross section, scale bars are 100nm and 5nm respectively. Denoted regions are (1) GaSb buffer, (2) bottom AlSb barrier, (3) GaSb hole well, (4) InAs electron well and (5) AlSb top barrier.

Materials were processed into 6-lead Hall bars using standard photolithography techniques. 200 μm long, 20 μm wide Hall bar mesas with Hall voltage lead spacings of 100 μm were etched using a citric-acid based solution. 500 Å Ti/3000 Å Au metal layers were e-beam evaporated for Ohmic contacts to the electron-hole layers. No top gate dielectric or passivation was applied in the reported devices. A top view micrograph of an actual device is shown in the inset of Fig3. Processed devices were cooled down to ~1.8 K in a Lakeshore 9709A Hall system and electrical measurements were performed using low frequency (9Hz) lock-in technique with a time constant of 30 ms. The Hall bar's drain/source were connected in series with a 25 kΩ resistor load under a 10 mV AC bias. The measurements were performed under a DC bias range from -2.5 to 1V applied to the backgate and a magnetic field sweep from 0 to 8T.

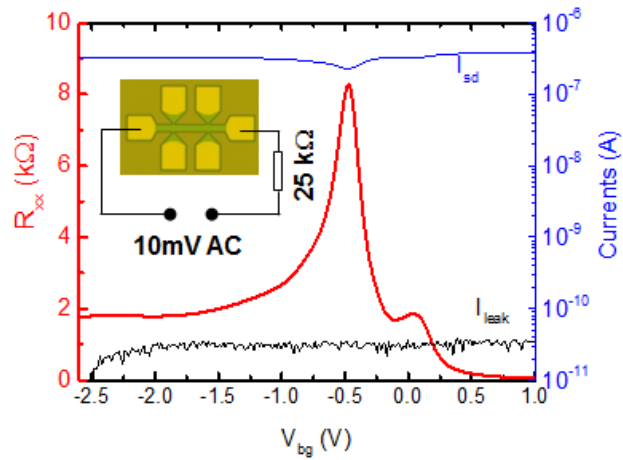

Fig3- Gate bias dependent characteristics of the Hall bar device. Inset: top view micrograph of a hall bar in the schematic circuit.

Fig 3 shows the transfer curve $I_{sd}$ vs. $V_{bg}$ at zero magnetic field under -2.5 to 1 V back gate bias. The $I_{sd}$ exhibits an ambipolar behavior with electron characteristics above -0.5 V, and hole characteristics below -0.5 V. The minimum current of 240nA is still 3 orders of magnitude larger than the gate leakage current shown in Fig3. The dip of source-drain current around -0.5V corresponds to the charge neutrality point (CNP), with an 8 kΩ peak longitudinal resistance that can attain 94 kΩ under applied magnetic field (not shown). At 0T, the ratio of peak resistance over saturated electron (hole) resistance at large positive (negative) gate bias is 100 (4). At V=-0.4V, just slightly above the charge neutrality point, the field dependent longitudinal resistance ($R_{xx}$ or magnetoresistance) (Fig4a) and transverse resistance ($R_{xy}$ or Hall resistance) (Fig 4b) exhibit clear quantum Hall effects with a Landau filling level ν=1 at B > 4.5T, which is a clear signature of high material quality. A broader perspective on the field-dependent resistances is shown in the 2D maps in Figs 4c and d. In Fig 4c, the peak magnetoresistance clearly shows Landau fans for both electrons and holes on the right and left side of the CNP (at $V_{bg}$=-0.5V), respectively. The oscillation fringes are visible even inside the CNP regime (horizontal fringes between -1V and 0V in Fig 4c), but the periodicity is less sensitive to back gate bias than it is in the electron and hole regimes, which indicates a constant sheet charge $N_s = \frac{e}{h\Delta(1/B)}$ where e is electron charge, h is the Plank constant and $\Delta(1/B)$ is the periodicity of the $R_{xx}$ oscillations in the reciprocal field. This can be explained by a residual electron channel that could not be modulated by the back gate when holes start to accumulate in the GaSb quantum well, screening the back gate action with respect to the electron channel in the InAs well. The presence of these residual electrons can also be seen in the sign of the Hall resistance $R_{xy}$ with the convention that positive $R_{xy}$ indicates electron-like and negative $R_{xy}$ indicates hole-like behaviors. At small field (B <1T), $R_{xy}$ stays positive even for large negative $V_{bg}$ due to larger electron mobility than hole mobility, and only flips sign at higher field where the contribution of mobility is weaker and the resistance is more dependent on the carrier concentration difference.[14]

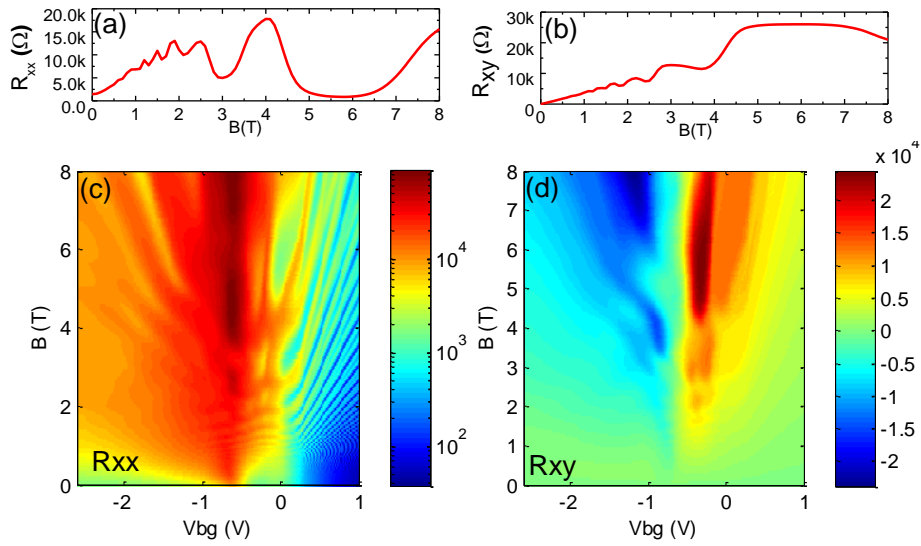

Fig4- Field dependent (a) magneto resistance and (b) Hall resistance at $V_{bg}$=-0.4 V. 2D maps of (c) magneto resistance and (d) Hall resistance for all fields and gate biases.

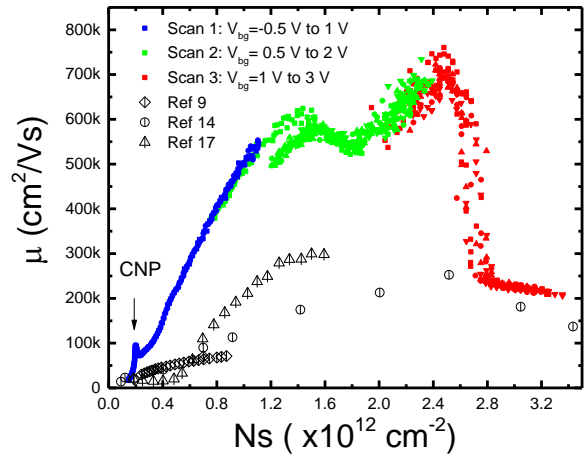

Fig5-Electron mobility vs. sheet charge density of InAs/GaSb double quantum well structures reported in this work and published data from Ref [9,14,17].

The field dependent magneto and hall resistances allow for extraction of carrier mobility and sheet charge density. In the hole regime, a two carrier fitting is necessary, but in the electron regime, the linearity of $R_{xy}$ vs. B (not shown) suggests that single field measurement is adequate. This enables us to extend the gate bias range to explore the detailed behavior of electrons in the InAs well. The mobility vs sheet charge density for three separate measurements with different gate bias ranges (-0.5V to 1V for blue dots, 0.5V to 2V for green dots and 1 to 3V for red dots) is shown in Fig 5, together with published data on the InAs/GaSb double quantum well system for comparison. From 0 to -0.5 V, the system is near CNP, the sheet charge remains relatively constant at $N_s$ ~$2\times10^{11}$ cm$^{-2}$ independent of gate bias, and the steep drop of carrier mobility is an indication of the CNP. Near the CNP, the mobility is almost 100,000 cm$^{-2}$/Vs, which is further proof of high material quality and is in line with the appearance of the Landau filling level $\nu$=1 at small field (4.5T) in Fig 4. As $V_{bg}$ increases positively from 0V, the mobility increases steadily with sheet charge, due to screening effects. The mobility reaches 500,000 cm$^2$/Vs at a sheet charge density of $8\times10^{11}$cm$^{-2}$ and peaks above 700,000 cm$^2$/Vs before dropping to ~200,000

cm$^2$/Vs at Ns= 2.5x10$^{12}$ cm$^{-2}$. The abrupt drop, which was also observed by Knez [14], is likely due to filling of a higher energy level with lower mobility.[34] At Ns < 1x10$^{12}$cm$^{-2}$, the linearity between the mobility and sheet charge density suggests that the mobility is limited by interface scattering,[14,24] thus justifying the higher mobility in our devices grown on GaSb substrates in comparison with those grown on GaAs substrates that suffer from rougher interfaces. The sheet charge concentration varies linearly with gate bias with slope of 7.5x10$^{11}$ cm$^{-2}$/V, which is significantly more efficient than back gate modulation of 0.4x10$^{11}$ cm$^{-2}$/V in Ref [25]. This value is also higher than top gate modulation of 1.5 x10$^{11}$ cm$^{-2}$/V in Ref [25] and Ref [17].

In summary, we demonstrate high material quality, high mobility InAs/GaSb quantum well structures grown on GaSb substrate. The backgate configuration has been shown to efficiently control the conductance channels through the use of a low leakage AlSb barrier as the backgate dielectric. The charge neutrality region is clearly revealed over an 8T magnetic field sweep and a 3.5V backgate bias range. The geometry of the backgate will facilitate realization of complicated device structures including Corbino disks, which remain very challenging if done with top gate configuration. When combined with a top gate, independent, efficient control of both electron and hole channels will allow exploration of the complete phase diagram of this InAs/GaSb inverted quantum well structure[12].

The authors wish to thank Prof. Leo Kouwenhoven from Delft University of Technology, Netherlands, Prof. Charles Marcus from Niels Bohr Institute, Denmark, Prof. Michael Manfra from Purdue University, USA, Prof. Vlad Pribiag at the University of Minnesota, USA for insightful suggestions and encouragement. We would also like to acknowledge Drs. Gunjana Sharma, Elias Flores, Marcel Musni, Mark Gyure, Andrey Kiselev, Steven Bui, John Yeah, Rajesh Rajavel, Tahir Hussain and David Chow at HRL labs for technical support and fruitful discussions. The authors are grateful for support from Microsoft Station Q.

**List of reference:**